\documentclass[prd,twocolumn,showpacs,preprintnumbers,amsmath,amssymb]{revtex4}
\usepackage{graphicx}
\usepackage{epsfig}
\usepackage{dcolumn}
\usepackage{bm}

\newcommand{\pspp}{\psi(3770)}
\newcommand{\psip}{\psi(2S)}
\newcommand{\psp}{\psi(2S)}
\newcommand{\jpsi}{J/\psi}

\newcommand{\chicJ}{\chi_{cJ}}
\newcommand{\chicz}{\chi_{c0}}
\newcommand{\chico}{\chi_{c1}}
\newcommand{\chict}{\chi_{c2}}
\newcommand{\EE}{e^+e^-}
\newcommand{\MM}{\mu^+\mu^-}

\newcommand{\pp}{\pi^+\pi^-}

\newcommand{\jpsito}{J/\psi \rightarrow }
\newcommand{\psipto}{\psp \rightarrow }
\newcommand{\pspto}{\psp \rightarrow }

\newcommand{\chiczto}{\chi_{c0} \rightarrow }

\newcommand{\chictto}{\chi_{c2} \rightarrow }
\newcommand{\bfg}{\begin{figure}}
\newcommand{\efg}{\end{figure}}
\newcommand{\bitm}{\begin{itemize}}
\newcommand{\eitm}{\end{itemize}}
\newcommand{\bnum}{\begin{enumerate}}
\newcommand{\enum}{\end{enumerate}}
\newcommand{\btbl}{\begin{table}}
\newcommand{\etbl}{\end{table}}
\newcommand{\btbu}{\begin{tabular}}
\newcommand{\etbu}{\end{tabular}}

\newcommand{\PP}{\pi^+\pi^-}
\newcommand{\KK}{K^+K^-}

\newcommand{\beq}{\begin{equation}}
\newcommand{\edq}{\end{equation}}
\def\thg{\theta_{\gamma}}
\def\thm{\theta_{M}}
\def\phm{\phi_{M}}

\begin{document}
\title{\boldmath  Measurement of the $\chict$ Polarization in $\pspto \gamma \chict$}
\author{ M.~Ablikim$^{1}$, J.~Z.~Bai$^{1}$, Y.~Ban$^{10}$,
  J.~G.~Bian$^{1}$, X.~Cai$^{1}$, J.~F.~Chang$^{1}$,
  H.~F.~Chen$^{16}$, H.~S.~Chen$^{1}$, H.~X.~Chen$^{1}$,
  J.~C.~Chen$^{1}$
% ...
  \\(BES Collaboration)\\
} \affiliation{
  $^1$ Institute of High Energy Physics, Beijing 100039, People's Republic of China\\
% ...
$^{10}$ Peking University, Beijing 100871, People's Republic of China\\
% ...
$^{16}$ University of Science and Technology of China, Hefei 230026, People's Republic of China\\
% ...
} \date{\today}

\begin{abstract}
The polarization of the $\chict$ produced in $\psip$ decays into
$\gamma \chict$ is measured using a sample of $14 \times 10^6$ $\psip$
events collected by BESII at the BEPC. A fit to the $\chict$
production and decay angular distributions in $\pspto \gamma \chict$,
$\chictto \PP$ and $\KK$ yields values $x=A_1/A_0=2.08\pm0.44$ and
$y=A_2/A_0=3.03\pm0.66$, with a correlation $\rho=0.92$ between them,
where $A_{0,1,2}$ are the $\chict$ helicity amplitudes. The
measurement agrees with a pure $E1$ transition, and $M2$ and $E3$
contributions do not differ significantly from zero.
\end{abstract}


\maketitle

\section{Introduction}

The radiative transition between charmonium states has been studied
extensively by many authors~\cite{PRD42p2293, PRD21p203, PRD26p2295,
  PRD25p2938, PRD28p1692, PRD28p1132}. In general, it is believed that
$\pspto \gamma \chicJ$ is dominated by the $E1$ transition, but with
some $M2$ (for $\chico$ and $\chict$) and $E3$ (for $\chict$)
contributions due to the relativistic correction.  These contributions
have been used to explain the big differences between the calculated
pure $E1$ transition rates and the experimental
results~\cite{PRD21p203}.  They will also affect the angular
distribution of the radiative photon. Thus the measurement of the
angular distribution may be used to determine the contributions of the higher
multipoles in the transition.

Furthermore, for $\pspto \gamma \chict$, the $E3$ amplitude is
directly connected with $D$-state mixing in $\psp$ which has been
regarded as a possible explanation of the large leptonic annihilation
rate of $\pspp$~\cite{PRD28p1132}. Since recent
studies~\cite{wympspp,wymkskl} also suggest the $S$- and $D$-wave
mixing of $\psp$ and $\pspp$ may be the key to solve the longstanding
``$\rho\pi$ puzzle'' and to explain $\pspp$ non-$D\bar{D}$ decays, the
experimental information on multipole amplitudes gains renewed
interest.

Decay angular distributions were studied in $\pspto \gamma \chict$
by the Crystal Ball experiment using $\pspto\gamma\gamma
J/\psi$~\cite{Cball}; the contribution of the higher multipoles
were not found to be significant but the errors were large due to
the limited statistics. In the present analysis,
$\pspto\gamma\chict\to\gamma\PP$ or $\gamma\KK$ decays will be
used for a similar study. The analysis on these channels has the
advantage that there is no background from $\chico$ since the
$\chico\to\PP$ and $\KK$ processes are forbidden by parity
conservation.

\section{The BES Experiment}

The data used for this analysis are taken with the BESII detector
at the BEPC storage ring operating at the $\psp$. The number of
$\psp$ events is $14.0 \pm 0.6$ million~\cite{moxh}, determined
from the number of inclusive hadrons.

The Beijing Spectrometer (BES) detector is a conventional
solenoidal magnet detector that is described in detail in
Ref.~\cite{bes}; BESII is the upgraded version of the BES
detector~\cite{bes2}. A 12-layer vertex chamber (VC) surrounding
the beam pipe provides trigger information. A forty-layer main
drift chamber (MDC), located radially outside the VC, provides
trajectory and energy loss ($dE/dx$) information for charged
tracks over $85\%$ of the total solid angle. The momentum
resolution is $\sigma _p/p = 0.017 \sqrt{1+p^2}$ ($p$ in
$\hbox{\rm GeV}/c$), and the $dE/dx$ resolution for hadron tracks
is $\sim 8\%$.  An array of 48 scintillation counters surrounding
the MDC measures the time-of-flight (TOF) of charged tracks with a
resolution of $\sim 200$ ps for hadrons.  Radially outside the TOF
system is a 12 radiation length, lead-gas barrel shower counter
(BSC).  This measures the energies of electrons and photons over
$\sim 80\%$ of the total solid angle with an energy resolution of
$\sigma_E/E=22\%/\sqrt{E}$ ($E$ in GeV).  Outside of the
solenoidal coil, which provides a 0.4~Tesla magnetic field over
the tracking volume, is an iron flux return that is instrumented
with three double layers of counters that identify muons of
momentum greater than 0.5~GeV/$c$.

A GEANT3 based Monte Carlo (MC) program with detailed consideration of
detector performance (such as dead electronic channels) is used to
simulate the BESII detector.  The consistency between data and Monte
Carlo has been carefully checked in many high purity physics channels,
and the agreement is quite reasonable.

MC samples of $\psipto\gamma\chi_{c0,2}\to \gamma\PP$ and
$\psipto\gamma\chi_{c0,2}\to \gamma\KK$ are generated according to
phase space to determine normalization factors in the partial wave
analysis.  MC samples of $\EE\to (\gamma)\EE$, $\psipto(\gamma)\EE$,
$\EE\to(\gamma)\MM$, $\psipto(\gamma) \MM$, and $\psipto X \jpsi$,
$\jpsito (\gamma) \MM$ ($X \to$ $\gamma\gamma$, $\pi^0\pi^0$, $\PP$,
and $\eta$) are used for background estimation.

\section{Event Selection}\label{evtslc}

For the decay channels of interest, there are two high momentum
charged tracks and one low energy photon. The candidate events are
required to satisfy the following selection criteria:

\begin{enumerate}

\item At least one photon candidate is required.  A neutral cluster is
  considered to be a photon candidate when the angle between the
  nearest charged track and the cluster in the $xy$ plane is greater
  than $15^{\circ}$, the first hit is in the beginning six radiation
  lengths of the BSC, and the angle between the cluster development direction in
  the BSC and the photon emission direction in the $xy$ plane is less
  than $37^{\circ}$. There is no restriction on the number of extra
  photons.
\item Two good charged tracks with net charge zero are required.  Both
  tracks must satisfy $|\cos\theta|<0.65$ , where $\theta$ is the
  polar angle of the track in the laboratory system. This angular
  region allows use of the $\mu$ counter information to eliminate
  $\MM$ background.
 \item To remove Bhabha events, the total energy deposited in the BSC
   energy by the two charged tracks is required to be less than 1~GeV,
   or $\chi^{dE/dx}_e = \frac{(dE/dx)_{\rm meas} -(dE/dx)_{\rm exp}}{
   \sigma}$ for each track is required to be less than -3. Here
   $(dE/dx)_{\rm meas}$ and $(dE/dx)_{\rm exp}$ are the measured and
   expected $dE/dx$ energy losses for electrons, respectively, and
   $\sigma$ is the experimental $dE/dx$ resolution.  This removes
   almost all events with two electron tracks but keeps the efficiency
   high for the signal channels.
 \item To remove $\MM$ backgrounds, $MUID^++MUID^-<3$ in the
   $\gamma\PP$ channel and $MUID^++MUID^-<5$ in the
   $\gamma\KK$ channel are required.  Here $MUID$ is the number of
   $\mu$ counter hits matched with the MDC track and ranges from 0 to
   3.  ``0" means not a $\mu$ track, while ``1'', ``2'' and ``3''
   means a loose, medium, or strong $\mu$ candidate~\cite{muid}.
 \item To remove cosmic rays, $|t_{TOF}^+-t_{TOF}^-|<4$~ns is
required, where $t_{TOF}$ is the time recorded by the TOF.
This removes all the cosmic ray events with almost 100\%
efficiency for the channels of interest.
 \item Four-constraint kinematic fits are performed with the two charged
tracks and the photon candidate with the largest BSC energy under the
hypotheses that the
two charged tracks are either $\PP$ or $\KK$, and  the kinematic
chisquares, $\chi^2_{\pi}$ and $\chi^2_{K}$, are determined.  If
$\chi^2_{\pi}<\chi^2_{K}$ and the confidence level of the fit to
$\pspto \gamma \PP$ is greater than 1\%, the event is categorized
as $\gamma\PP$; otherwise, if $\chi^2_{K}<\chi^2_{\pi}$ and the
confidence level of the fit to $\pspto \gamma \KK$ is greater than
1\%, the event is categorized as $\gamma\KK$.

\end{enumerate}

\begin{figure*}[htbp]
  \centering \includegraphics[width=15.5cm,angle=0]{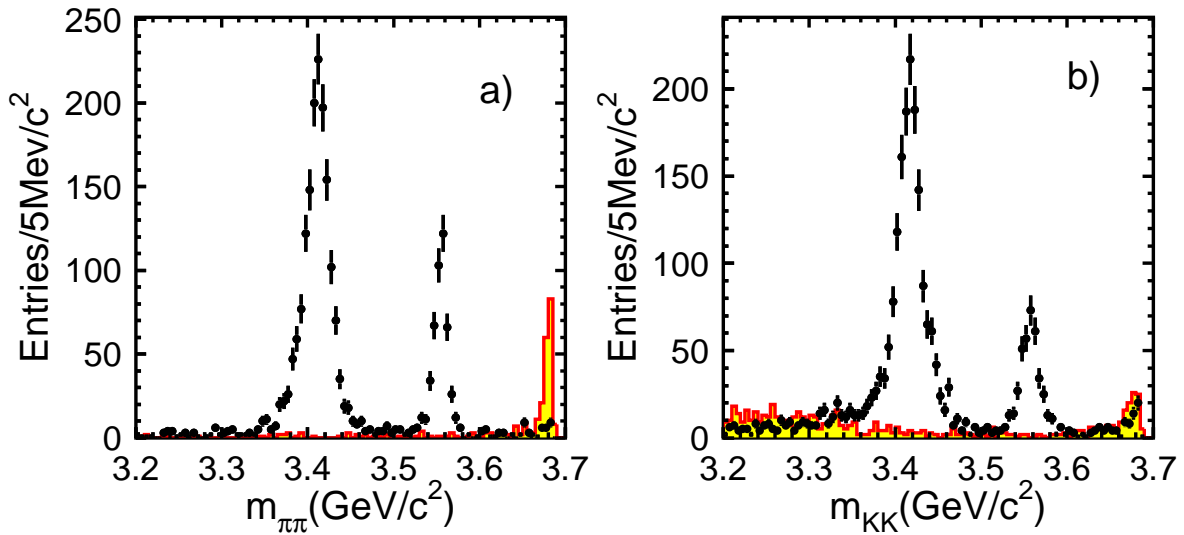}
  \caption{\label{sig_bkg}Invariant mass distributions of the two
charged tracks in (a) $\gamma\PP$ and (b) $\gamma\KK$. Dots
with error bars are data, and the shaded histograms are the MC
simulated backgrounds.}
\end{figure*}

After imposing the above requirements, the invariant mass distributions
for the selected $\gamma\pp$ and $\gamma\KK$ candidates are shown in
Fig.~\ref{sig_bkg}. Clear $\chicz$ and $\chict$ signals can be
seen while the background level is low.

Simulated background events passing the selection criteria for
the $\gamma\pp$ and $\gamma\KK$ channels are also plotted in
Fig.~\ref{sig_bkg}. The excess background in the $\gamma\PP$ mode near
3.7~GeV/$c^2$ is due to the large $\pspto \PP$ branching ratio from
the PDG~\cite{pdg}. The backgrounds under the signal regions are
$\MM(\gamma)$ and $\EE(\gamma)$ events either from QED processes or from
$\psp$ decays.

Requiring the invariant mass of the two charged tracks be between
$3.54$ and $3.57$~GeV/$c^2$ to select $\chict$, 418 $\gamma\PP$
events and 303 $\gamma\KK$ events are selected. The fractions of
background are $(1.6\pm 0.5)\%$ for $\gamma\PP$ and $(2.8\pm
0.6)\%$ for $\gamma\KK$, as estimated from Monte Carlo simulation,
in agreement with the expectation from the measured
misidentification efficiencies in data.

Monte Carlo simulation also determines that the $\KK$
contamination in the $\PP$ sample is about 9\%, and the $\PP$
contamination in the $\KK$ sample is about 34\%. The effect of the
cross contamination on the fit of the helicity amplitudes will be
discussed later.

\section{The Fit of the Helicity Amplitudes}


The $\psip\to\gamma\chict$ helicity amplitudes are determined
by a maximum likelihood fit to the decay angular
distribution~\cite{3161, 3065}
\begin{eqnarray}
\label{fitform}
 W_2(\thg,\thm,\phm)={ 3x^2\sin^2\thg \sin^2
2\thm}+  \\ \nonumber
{(1+\cos^2\thg)[(3\cos^2\thm-1)^2+{\frac{3}{2}}y^2\sin^4\thm]}+\\
\nonumber {\sqrt{3}x\sin2\thg\sin2\thm[3\cos^2\thm-1}-\\ \nonumber
{{\frac{1}{2}\sqrt{6}}y\sin^2\thm]\cos\phm} \\ \nonumber
{+\sqrt{6}y\sin^2\thg\sin^2\thm(3\cos^2\thm-1)\cos2\phm},
\end{eqnarray}
where $x=A_1/A_0$, $y=A_2/A_0$, $A_{0,1,2}$ are the
$\chi_{c2}$ helicity amplitudes, $\thg$ is the polar angle of the
photon in the laboratory system, and $\thm$ and $\phm$ are the polar
and azimuthal angles of one of the mesons in the $\chict$ rest
frame with respect to the $\gamma$ direction. $\phm=0$ is defined
by the electron beam direction.

\begin{figure}[htbp]
  \centering \includegraphics[width=8cm,angle=0]{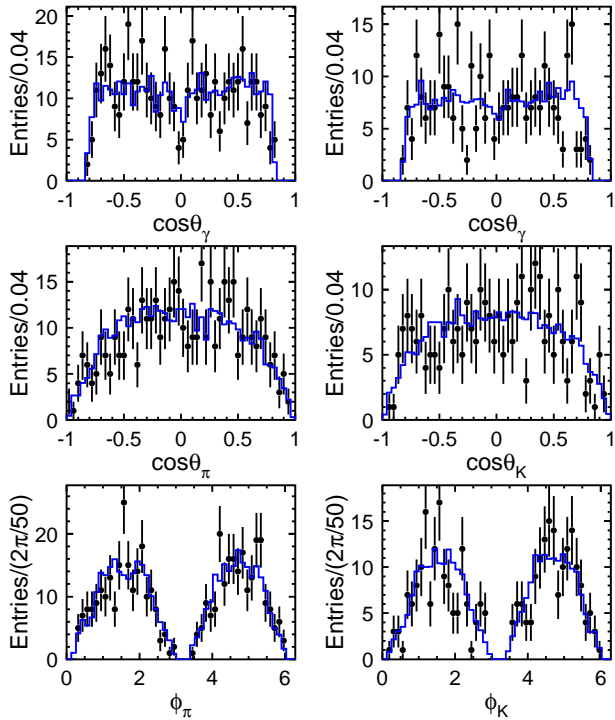}
  \caption{\label{compdtmc}Comparison between data and the final fit for
    $\gamma\PP$(left) and $\gamma\KK$ (right), where dots with
    error bars are data and the histograms are the fit.}
\end{figure}

Fitting the $\gamma\PP$ and $\gamma\KK$ data, we
obtain
\begin{eqnarray*}
 x_{\pi}&=&1.97\pm 0.64, \,\, y_{\pi}=3.03\pm 1.07,\,\,
 \rho_{\pi}=0.96, \\
 x_{K}&=&1.77\pm 0.54,\,\, y_{K}=2.36\pm 0.82,\,\,\rho_{K}=0.94,
\end{eqnarray*} where the errors are statistical and $\rho_{\pi}$,
$\rho_{K}$ are the correlation factors between $x$ and $y$ for
$\gamma\PP$ and $\gamma\KK$, respectively. The comparison between
data and the fit is shown in Fig.~\ref{compdtmc}. Good agreement
is observed in all angular distributions for both the $\gamma\PP$
and $\gamma \KK$ channels.

Since the value of the likelihood function does not provide a
measurement of the goodness of fit, Pearson's $\chi^2$ test is
used. The data are divided into
$3\times 3\times 4=36$ bins in $\cos\thg$, $\cos\thm$ and $\phm$.
The $\chi^2$ is calculated using
$$
\chi^2=\sum\limits_i\frac{(n_i^{DT}-n_i^{MC})^2}{n_i^{DT}},
$$ where $n_i^{DT}$ is the observed number of events in the $i$th bin
and $n_i^{MC}$ is the corresponding number of events predicted by
Monte Carlo using $x$ and $y$ fixed to the values determined in
this analysis. We obtain $\chi_{\pi}^2/ndf=30.19/35=0.86$ and
$\chi_{K}^2/ndf=43.57/35=1.24$ for the $\gamma\PP$ and $\gamma\KK$
channels, respectively, where $ndf$ is the number of the degree of
freedom. These results show that the fits are good.

\section{Error analysis}

\subsection{Input output checking}

The fitting procedure is tested using Monte Carlo simulated
samples.
With input parameters $x_{in}=\sqrt{3}\approx 1.732$ and
$y_{in}=\sqrt{6}\approx 2.449$, fitting a Monte Carlo sample of 50,000
selected events gives the results $x_{out}=1.74\pm0.04$,
$y_{out}=2.45\pm0.07$, and $\rho=0.94$, which are in good agreement
with the input values, indicating the validity of the fitting
procedure.

\begin{figure}[htbp]
  \centering \includegraphics[width=8cm,height=8cm,angle=0]{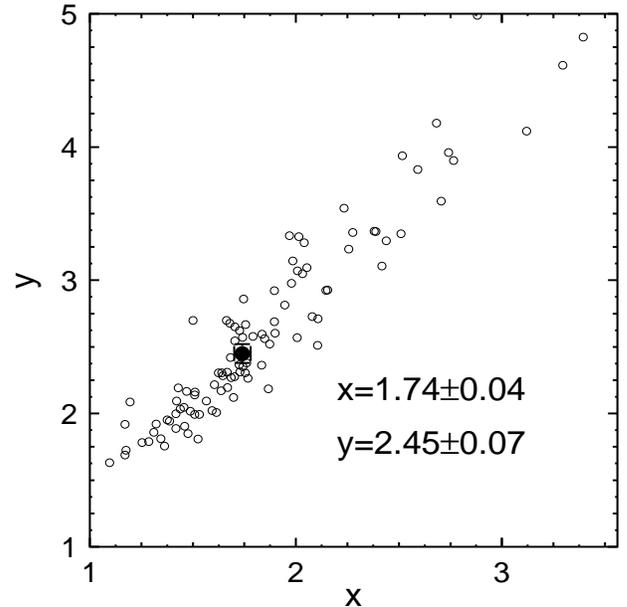}
  \caption{\label{xydis_p} Distribution of fitting results for $x$ and
$y$ for Monte Carlo simulated samples. The black dot with error bar is
the result for all 50,000 events. The circles are the fitting results
for the subsets after dividing the sample into 100 subsets.}
 \end{figure}

Dividing the 50,000 events into 100 subsets of 500 events each
(about the same size as the real data sample) gives the distribution
of fitting results shown in Fig.~\ref{xydis_p}.  $x$ and $y$ are
positively correlated, and the fitting results are distributed in a
relatively broad area due to the limited statistics of the subsets.

\subsection{Systematic errors}

Systematic errors from background,
from the $\gamma\PP$ and $\gamma\KK$ cross contamination, from the
Monte Carlo simulation of the detector response, etc. are
considered.

\subsubsection{Background contamination}

Backgrounds remaining after event selection are $\MM(\gamma)$ and
$\EE(\gamma)$ events, and the fractions of backgrounds in $\gamma\PP$
and $\gamma\KK$ channels are estimated by Monte Carlo simulation and
checked with data. In the fit, background is not considered, but the
effect on the helicity amplitudes is estimated using Monte Carlo
simulation.  By adding the amount of MC background mentioned in
Sec.~\ref{evtslc} into the pure MC sample, the fit yields shifts of
the fit results. These shifts are taken as corrections to the results
obtained from data. By varying the background fraction in the fit, the
uncertainty due to the background contamination can also be determined.
It is found that the corrections to the $\gamma\PP$ results are
$\Delta_{\pi x}=0.19\pm 0.04$, $\Delta_{\pi y}=0.34\pm 0.07$; and for
$\gamma\KK$, $\Delta_{K x}=0.25\pm 0.04$, $\Delta_{K y}=0.47\pm 0.08$.

\subsubsection{$\gamma\PP$ and $\gamma\KK$ cross contamination}

In order to study the error from $\gamma\PP$ and $\gamma\KK$ cross
contamination, Monte Carlo samples of $\gamma\PP$ and $\gamma\KK$ with
$x=\sqrt{3}$, $y=\sqrt{6}$ are generated and mixed according to the
amount of cross contamination determined by Monte Carlo simulation. It
is found that the results from this mixed sample are mostly unchanged from
those of the pure Monte Carlo sample, even when the contamination is
doubled. This is understandable since the angular distributions of
$\gamma\PP$ and $\gamma\KK$ are identical. From the comparisons of
many Monte Carlo samples with different fractions of cross
contamination, the errors on $x$ and $y$ are determined to be $0.01$
and $0.06$ for $x_{\pi}$ and $y_{\pi}$, and $0.10$ and $0.14$ for
$x_{K}$ and $y_{K}$, respectively.

\subsubsection{MC simulation of the detector response}

\begin{figure}[htbp]
  \centering
        \includegraphics[width=8cm]{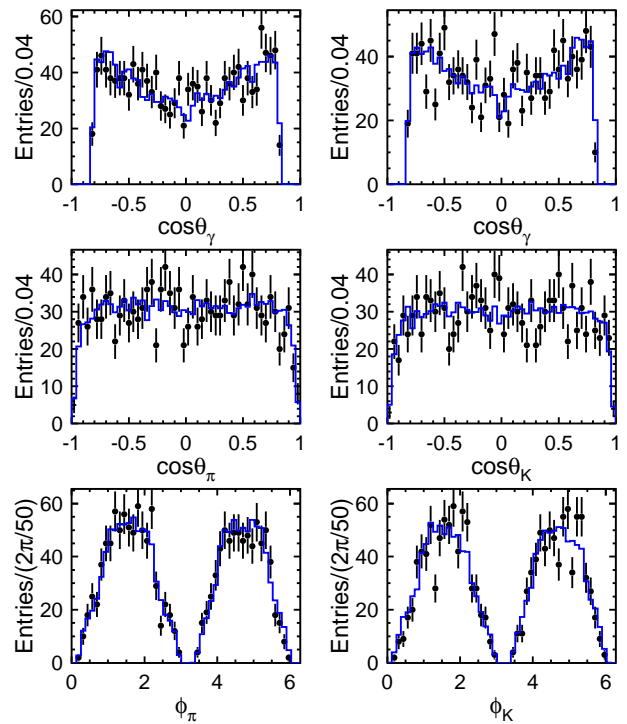}
  \caption{Comparison of angular distributions between data (dots with
    error bars) and Monte Carlo simulation (histograms) for
    $\chiczto\PP$ (left) and $\chiczto\KK$ (right). Fitting the
    $\chi_{c0}$ angular distributions provides a way to estimate the
    systematic error due to the Monte Carlo simulation of the detector
    response.}
\label{fig:chicz}
\end{figure}

The consistency between data and the Monte Carlo simulation of the
detector response for $\chict$ events can be determined using
$\chicz$ events, although the absolute angular distributions are
different. The angular distribution of $\chicz$ decays is
unambiguous, i.e. $ W_{0}=1+\cos^2\theta_{\gamma}$. Note that
Eq.~1 with the $(3 \cos^2 \theta_M-1)$ term replaced by 1 is equal
to $W_{0}$ when both $x$ and $y$ are zero. Therefore, if we fit
the angular distribution of $\chicz$ the same as $\chict$ using
the modified Eq.~\ref{fitform}, $x$ and $y$ should be 0. The
difference from zero gives the systematic error due to the MC
simulation of the detector response. For the $\gamma\PP$ channel,
0.18, 0.05, and 0.24 are obtained for $x_{\pi}$, $y_{\pi}$ and
$\rho_{\pi}$, respectively, and for the $\gamma\KK$ channel, 0.13,
0.08, and -0.24 are obtained for $x_{K}$, $y_{K}$ and $\rho_{K}$.
The results are dominated by the statistical errors of the fit due
to the limited $\chicz$ samples, although they are already much
larger than the corresponding $\chict$ samples. Comparisons
between data and Monte Carlo simulation for $\chicz$ events are
shown in Fig.~\ref{fig:chicz}; good agreement is observed.

\subsubsection{Other sources}

Other sources of error are from systematic errors associated with the
simulation of the mass resolution of the $\chict$, the photon
detection efficiency, the MDC tracking efficiency, the kinematic fit,
the total number of the $\psp$ events, the trigger efficiency,
etc. These systematic errors will affect a branching ratio
measurement, but will not affect the measurement of the angular
distribution. Their effects on the helicity amplitude measurements are
neglected.

\subsubsection{Total systematic error}

The systematic errors and the correlation factors from all the
above sources are listed in Table~\ref{systxy}. Here the
correlation factors ($\rho_{\pi}$ and $\rho_{K}$) from background
contamination and $\gamma\PP$ and $\gamma\KK$ cross contamination
are set to $1$, and the total correlation factor $\rho$ is
calculated with
$\rho=\Sigma_i\frac{\rho_i\sigma_{xi}\sigma_{yi}}{\sigma_{x}\sigma_{y}}$,
where $i$ runs over all the systematic errors. The
total systematic errors are $0.19$ and $0.11$ for $x$ and $y$ in
$\gamma\PP$ and $0.17$ and $0.18$ in $\gamma\KK$.

\begin{table}[htbp]
\caption{Summary of the systematic errors and correlations.}
\begin{center}
\begin{tabular}{c|ccc|ccc}\hline
 Source                      &$x_{\pi}$ &$y_{\pi}$  &$\rho_{\pi}$  &$x_{K}$ &$y_{K}$  &$\rho_{K}$ \\\hline
Background contamination     &$0.04$    &$0.07$     &1             &$0.04$  &$0.08$   &1          \\
$\pi/K$ cross contamination  &$0.01$    &$0.06$     &1             &$0.10$  &$0.14$   &1          \\
MC simulation                &$0.18$    &$0.05$     &0.24          &$0.13$  &$0.08$   &-0.24      \\\hline
Total                        &$0.19$    &$0.11$     &0.29          &$0.17$  &$0.18$   &0.49       \\\hline
\end{tabular}
\end{center}
\label{systxy}
\end{table}

\section{Results and Discussion}\label{sec:final}


After applying the corrections due to the background
contamination, we obtain
\begin{eqnarray*}
x_{\pi}&=&2.16\pm0.64\pm0.19, \\
y_{\pi}&=&3.37\pm1.07\pm0.11,\\
\rho_{\pi}^{stat}&=&0.96, \,\,\,\,\rho_{\pi}^{sys}=0.29
\end{eqnarray*}
from $\gamma\PP$ and
\begin{eqnarray*}
x_{K}&=&2.02\pm0.54\pm0.17, \\
y_{K}&=&2.83\pm0.82\pm0.18, \\
\rho_K^{stat}&=&0.94, \,\,\,\,\rho_K^{sys}=0.49
\end{eqnarray*}
from $\gamma\KK$, where the first errors are statistical and the
second are systematic, and $\rho^{stat}$ and $\rho^{sys}$ are the
correlation factors between $x$ and $y$ of the statistical and
systematic errors.

Combining the statistical and systematic errors yields:
\begin{eqnarray*}
 x_{\pi}&=&2.16\pm0.67,\,\,
 y_{\pi}=3.37\pm1.08,\,\,\rho_{\pi}=0.93,\\
 x_{K}&=&2.02\pm0.57,\,\, y_{K}=2.83\pm0.84,\,\, \rho_{K}=0.91.
\end{eqnarray*}
The results from $\gamma\PP$ and $\gamma\KK$ are in good
agreement. Combining them, we obtain
\begin{eqnarray*}
   x&= &2.08 \pm 0.44, \\
   y&= &3.03 \pm 0.66, \\
   \rho&= &0.92.
\end{eqnarray*}
The combination assumes no correlation between $\gamma\PP$ and
$\gamma\KK$ for both statistical and systematic errors.


Comparing with the measurement obtained by the Crystal
Ball~\cite{Cball}, $a_2'(\psi'\to\gamma\chi)=0.132^{+0.098}_{-0.075}$,
this measurement gives the quadrapole amplitude
$a_2'=-0.051^{+0.054}_{-0.036}$ and the octupole amplitude
$a_3'=-0.027^{+0.043}_{-0.029}$ \cite{amplitudes}. Neither result significantly differs
from zero. The results are in good agreement with what is expected for
a pure $E1$ transition.
As for the D-state mixing of $\psip$, our results do not
contradict the previous theoretical calculation within one
standard deviation~\cite{PRD30p1924}.

\section{Summary}

The helicity amplitudes of $\pspto\gamma\chict$ are measured for
$\chictto \PP$ and $\KK$, and $x=2.08\pm 0.44$, $y=3.03\pm 0.66$ with
correlation $\rho=0.92$ are obtained. The results are in good
agreement with a pure $E1$ transition, but still do not have the
precision to strongly limit the higher multipoles.

\acknowledgments

The BES collaboration thanks the staff of BEPC for their hard
efforts and the members of IHEP computing center for their helpful
assistance. This work is supported in part by the National Natural
Science Foundation of China under contracts Nos. 19991480,
10225524, 10225525, the Chinese Academy of Sciences under contract
No. KJ 95T-03, the 100 Talents Program of CAS under Contract Nos.
U-11, U-24, U-25, and the Knowledge Innovation Project of CAS
under Contract Nos. U-602, U-34 (IHEP); by the National Natural
Science Foundation of China under Contract No. 10175060 (USTC), and
No. 10225522 (Tsinghua University); and by the Department of
Energy under Contract No. DE-FG03-94ER40833 (U Hawaii).

\end{document}